\documentclass[
]{ceurart}

\usepackage{listings}
\usepackage{svg}
\usepackage{graphicx}
\usepackage{sepfootnotes}
\usepackage{csquotes}
\usepackage{subfig}
\usepackage{listings}
\usepackage{xcolor}
\usepackage{hhline}

\usepackage[dvipsnames,svgnames]{xcolor} % text color
\usepackage[normalem]{ulem} % wavy underlines

\makeatletter
\font\uwavefont=lasyb10 scaled 700
\def\spelling{\bgroup\markoverwith{\lower3.5\p@\hbox{\uwavefont\textcolor{Red}{\char58}}}\ULon}
\def\grammar{\bgroup\markoverwith{\lower3.5\p@\hbox{\uwavefont\textcolor{LimeGreen}{\char58}}}\ULon}
\def\phrasing{\bgroup\markoverwith{\lower3.5\p@\hbox{\uwavefont\textcolor{RoyalBlue}{\char58}}}\ULon}

\newcommand\remove{\bgroup\markoverwith{\textcolor{red}{\rule[0.5ex]{2pt}{0.4pt}}}\ULon}
\newcommand\insertion{\bgroup\markoverwith{\textcolor{Green}{\rule[-0.5ex]{2pt}{0.6pt}}}\ULon}
\makeatother

\begin{document}

%%
%% Rights management information.
%% CC-BY is default license.
\copyrightyear{2024}
\copyrightclause{Copyright for this paper by its authors.
  Use permitted under Creative Commons License Attribution 4.0
  International (CC BY 4.0).}

%%
%% This command is for the conference information
\conference{The ISWC 2024 Posters Track}

%%
%% The "title" command
\title{Optimizing Traversal Queries of Sensor Data Using a Rule-Based Reachability Approach}

%%
%% The "author" command and its associated commands are used to define
%% the authors and their affiliations.
\author[1]{Bryan-Elliott Tam}[%
email=bryanelliott.tam@ugent.be,
orcid=0000-0003-3467-9755
]
\cormark[1]

\author[1]{Ruben Taelman}[%
orcid=0000-0001-5118-256X,
email=ruben.taelman@ugent.be,
url=https://www.rubensworks.net,
]
\author[1]{Julián {Rojas Meléndez}}[%
email=JulianAndres.RojasMelendez@UGent.be,
orcid=0000-0002-6645-1264,
url=https://julianrojas.org
]
\author[1]{Pieter Colpaert}[%
orcid=0000-0001-6917-2167,
email=pieter.colpaert@ugent.be,
url=https://pietercolpaert.be
]

\cortext[1]{Corresponding author.}

\address[1]{IDLab,
Department of Electronics and Information Systems, Ghent University – imec}

%%
%% Keywords. The author(s) should pick words that accurately describe
%% the work being presented. Separate the keywords with commas.

%%
%% This command processes the author and affiliation and title
%% information and builds the first part of the formatted document.
\maketitle

\begin{abstract}
% <!-- Context -->
Link Traversal queries face challenges in completeness and long execution time due to the size of the web.
Reachability criteria define completeness by restricting the links followed by engines.
However, the number of links to dereference remains the bottleneck of the approach.
% <!-- need -->
Web environments often have structures exploitable by query engines to prune irrelevant sources.
Current criteria rely on using information from the query definition and predefined predicate.
However, it is difficult to use them to traverse environments where logical expressions indicate the location of resources.
% <!-- task -->
We propose to use a rule-based reachability criterion that captures logical statements expressed in hypermedia descriptions within linked data documents to prune irrelevant sources.
%<!-- object -->
In this poster paper, we show how the Comunica link traversal engine is modified to
take hints from a hypermedia control vocabulary, to prune irrelevant sources.
% <!-- findings -->
Our preliminary findings show that by using this strategy, the query engine can significantly reduce the number of HTTP requests 
and the query execution time without sacrificing the completeness of results.
% <!-- conclusion -->
Our work shows that the investigation of hypermedia controls in link pruning of traversal queries is a worthy effort for optimizing web queries of unindexed decentralized databases.
\end{abstract}

\section{Introduction}

\sepfootnotecontent{sf:treeSpec}{
    \href{https://treecg.github.io/specification/}{https://treecg.github.io/specification/}
}

\href{https://lod-cloud.net/#diagram}{The increasing amount of available Linked Data on the Web}~\cite{Ermilov2013} prompts the need for efficient query interfaces.
During a typical query execution in a SPARQL endpoint, the endpoint takes the whole query load and delivers the results to the client.
This paradigm can lead to high workloads, which are partly responsible for the historically low availability of SPARQL endpoints~\cite{aranda2013}.
Researchers and practitioners have made efforts to introduce alternative Linked Data publication methods that enable client's participation in the query execution process~\cite{Verborgh2016TriplePF}.
The goal of those methods is to lower server-side workloads while keeping fast query execution to the client~\cite{Azzam2021}.
The TREE hypermedia specification is an effort in that direction~\cite{ColpaertMaterializedTREE, lancker2021LDS}, that introduces the concept of domain-oriented fragmentation of large RDF datasets.
For example, in the case of periodic measurements of sensor data, a fragmentation can be made on the publication date of each data entity.
A fragment can be considered an RDF document published in a server.
TREE aims to describes dataset fragmentation in ways that enable clients to easily fetch query-relevant subsets.
The data within a fragment are bound by constraints expressed through hypermedia descriptions~\cite{thomasFieldingPhdThesis}.
Each fragment contains relations to other pages, and those relations contain the constraints of the data of every reachable fragment.
In this paper, we refer to those constraints as domain-specific expressions.
They can be expressions such as $?t > \text{2022-01-09T00:00:00.000000} \implies \text{ex:afterFirstSeptember}$ 
given that $?t$ is the date of publication of sensor data and the implication pertains to the location of the data respecting the constraint.
In English, the expression means ``the data produced by the sensors after the first of September are stored at \texttt{ex:afterFirstSeptember}.''
Because of the hyperlinked nature of the documents network, clients must traverse them to find the relevant data to answer their queries.
We propose to use Link Traversal Query Processing (LTQP)~\cite{Hartig2016} as a query mechanism to perform those queries.

LTQP starts by dereferencing a set of user-provided URLs~\cite{Hartig2016}.
From these dereferenced documents, links to other documents are dereferenced recursively and inserted in an internal data store.
LDQL~\cite{hartig2016Ldql} is a theoretical query language to define the traversal of LTQP queries.
However, LDQL is centered around nested regular expressions, thus, is not made to express the traversal of links based on domain-specific expressions
such as time relations.
The subweb specifications language (SWSL)~\cite{bogaerts_rulemlrr_2021}, allows data providers to define traversal paths concerning the information they publish.
Thus, given that the query engine trusts the data publisher it can adapt its traversal to follow the paths given by the specification.
Akin to LDQL, it is difficult with the SWSL to express traversal using domain-specific expressions, because its syntax is centered around the matching of triple patterns and not reasoning rules or evaluation of literals.
Furthermore, SWSL does not propose a mechanism for using the query or input from the user to impact the source selection process, unlike LDQL.
Given those limitations, we propose to return to the more abstract concept of reachability criteria~\cite{hartig2012},
to define a mechanism of traversal centered around rules.

In this paper, we propose to use a boolean solver as the main link pruning mechanism for a reachability criterion to traverse TREE documents.
The logical operators are defined by the \href{https://w3id.org/tree/specification/}{TREE specification}.~\sepfootnote{sf:treeSpec}
As a concrete use case, we consider the publication of (historical) sensor data.
An example query is presented in Figure~\ref{lst:system} along with the triples representing the link between two documents expressed using the TREE specification.

\begin{figure}[h]
    \begin{minipage}{0.50\textwidth}
        \centering
\begin{lstlisting}[language=,frame=single]
PREFIX xsd: <http://www.w3.org/2001/XMLSchema#> 
PREFIX saref: <https://saref.etsi.org/core/>
PREFIX dahcc: <https://dahcc.idlab.ugent.be/Ontology/Sensors/>

SELECT * WHERE {
  ?s saref:hasTimestamp ?t;
    saref:hasValue ?result;
    saref:measurementMadeBy ?sensor.
  ?sensor dahcc:analyseStateOf ?stateOf;
    saref:measuresProperty {:property}.
  FILTER(?t="2022-01-03T10:57:54.000000"^^xsd:dateTime)
}
\end{lstlisting}
    \end{minipage}
    \hspace{0.05\textwidth}
    \begin{minipage}{0.43\textwidth}
        \centering
\begin{lstlisting}[language=,frame=single]
@prefix tree: <https://w3id.org/tree#> .
@prefix xsd: <http://www.w3.org/2001/XMLSchema#> .
@prefix ex: <https://example.be/> .
@prefix saref: <https://saref.etsi.org/core/> .

<> tree:relation [
    a  tree:GreaterThanOrEqualToRelation ;
    tree:node <nextNode> ;
    tree:value "2022-01-03T09:47:59"^^xsd:dateTime ;
    tree:path saref:hasTimestamp 
] .
\end{lstlisting}
    \end{minipage}
    \caption{On the left, is a SPARQL query to get sensor measurements and information about the sensor.
    On the right, is the hypermedia description of the location and constraint of the next fragment located in \texttt{ex:nextNode}.
    The constraint describes publication times ($?t$) where $?t>= \text{2022-01-03T09:47:59.000000}$.}
        \label{lst:system}
    \vspace*{-0.90cm}
\end{figure}

\section{A Rule-Based Reachability Criterion}

\sepfootnotecontent{sf:opensSource}{
The implementation, the queries and the evaluation are available at the following links:\newline
\href{https://github.com/constraintAutomaton/comunica-feature-link-traversal/tree/feature/time-filtering-tree-sparqlee-implementation}{https://github.com/constraintAutomaton/comunica-feature-link-traversal/tree/feature/time-filtering-tree-sparqlee-implementation},
\href{https://github.com/TREEcg/TREE-Guided-Link-Traversal-Query-Processing-Evaluation/tree/main}{https://github.com/TREEcg/TREE-Guided-Link-Traversal-Query-Processing-Evaluation/tree/main}
}

\sepfootnotecontent{sf:predicateCriterion}{
The query engine will always follow \texttt{ex:nextNode} from expressions following the schema "\texttt{ex:currentNode tree:relation [tree:node ex:nextNode]}" regardless of the constraints.
}

\sepfootnotecontent{sf:reachabilityCriterion}{
    We use a simplified formalization to illustrate the source selection mechanism and to not introduce unnecessary concepts for the aim of this poster paper. 
}

Most research on LTQP is centered around query execution in Linked Open Data environments.
Given the pseudo-infinite number of documents on the Web, traversing over all documents is practically infeasible.
To define completeness, different reachability criteria~\cite{hartig2012} were introduced to allow the discrimination of links.
Recently, an alternative direction was designed where the query engine uses the structure from the data publisher to guide itself towards relevant data sources~\cite{taelman2023, verborgh2020}.

\begin{figure}[htbp]
    \centering
    \includegraphics[width=\linewidth]{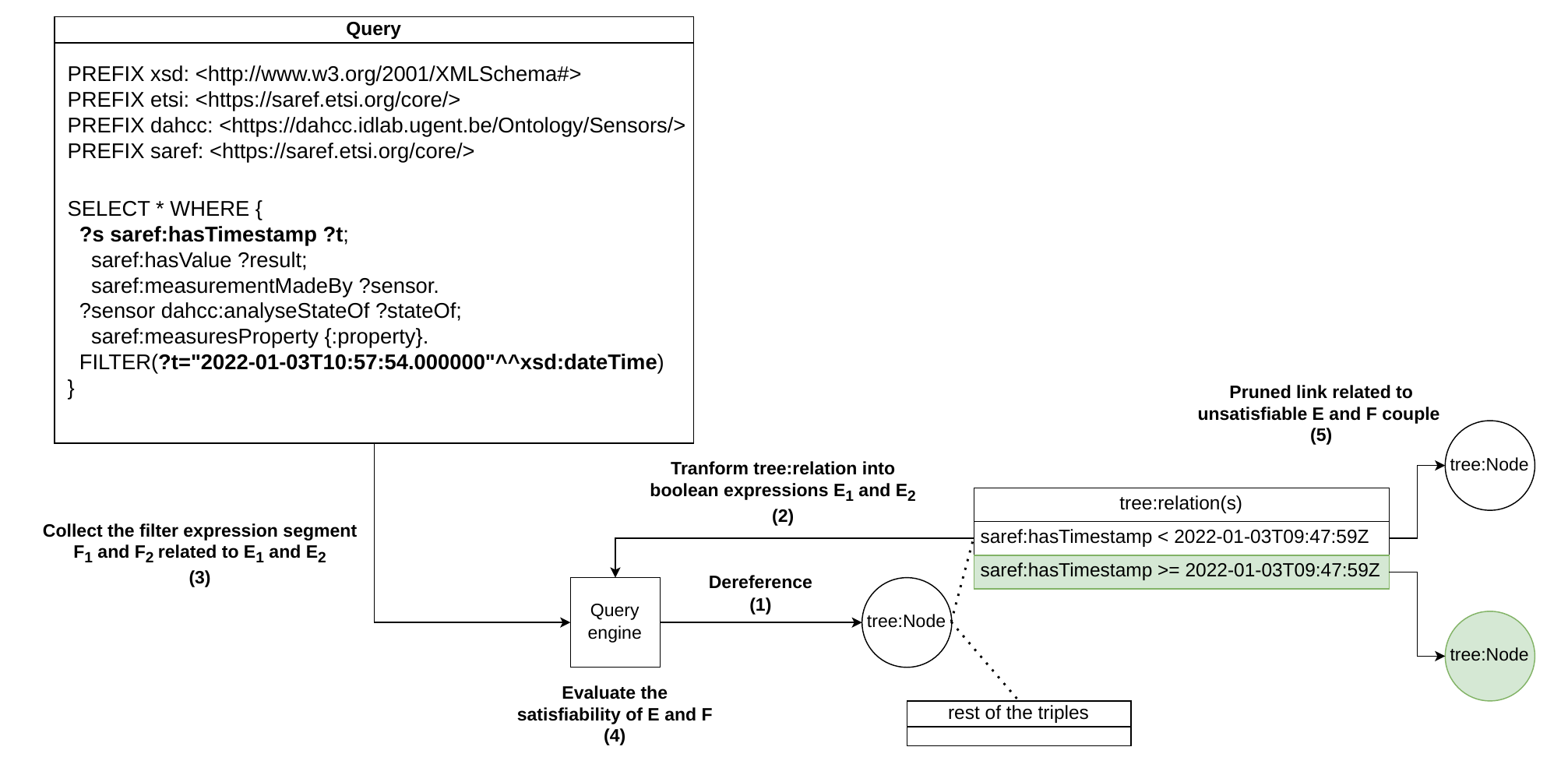}
    \caption{A schematization of our rule-based reachability criteria with a TREE document.
      First a TREE node is dereferenced, then the TREE relations are transformed into boolean expressions $E$,
      followed by the construction of $F$ from the filter expression related to the path of $E$ (the variable $t$ related to \texttt{saref:hasTimestamp}),
       then the satisfiability $E \land F$ is determined and finally links to non-query relevant data are pruned.}
    \label{fig:process}
  \end{figure}

We define our approach as a rule-based reachability criterion.
Our approach builds upon the concept of structural assumptions~\cite{taelman2023} to exploit the structural properties of TREE annotated datasets.
We therefore interpret the hypermedia descriptions of constraints in TREE fragments as boolean expressions $E$ ($?t>= \text{2022-01-03T09:47:59.000000}$ in Figure~\ref{lst:system}).
Upon discovery of a document, the query engine gathers the relevant triples to form the boolean expression of the constraint on the data of reachable fragments.
After the parsing of the expression, the filter expression $F$ of the SPARQL query is \textit{pushed down} into the engine's source selection component.
The source selection component can be formalized as a reachability criterion~\sepfootnote{sf:reachabilityCriterion} 
\begin{equation}
c(i) \rightarrow \{\mathrm{true}, \mathrm{false}\}
\end{equation}
where when it returns $\mathrm{true}$ the target IRI $i$ \emph{must} be dereferenced.
Finally, the two boolean expressions are evaluated to determine their satisfiability.
It can be formalized as determining if 
\begin{equation}
    c(i) = \exists x | (F(x) \land E_i) = \mathrm{true}
\end{equation}
hold true given $x$ is the variable targeted by $E_i$ and $i$ is the link towards the next fragment (\texttt{<nextNode>} from ``\texttt{<> tree:node <nextNode>}'' in Figure~\ref{lst:system}).
A variable targetted by $E$ is defined by an RDF object where the predicate as a value \texttt{?target} from the triple
defining the fragmentation path in the form ``''\texttt{?s tree:path ?target}'' (\texttt{saref:hasTimestamp} in Figure~\ref{lst:system}).
Upon satisfaction the IRI targeting the next fragment is added to the link queue otherwise the IRI is pruned.
The process is schematized in Figure~\ref{fig:process}.

\subsection{Preliminary Results}

We implemented our approach using the query engine Comunica~\cite{comunica}.
For evaluation, we executed four queries similar to the one in Figure \ref{lst:system}.~\sepfootnote{sf:opensSource}
They were executed over the DAHCC participant 31 dataset~\cite{dahcc_resource} (487 MB) with a timeout of two minutes.
We fragmented the dataset according to the TREE specification.
We use a B-tree topology with a depth of 1 using 100 and 1000 nodes ($n$).

\begin{table}[ht]
    \centering
    \begin{tabular}{|c|c|c|c|c|c|}
        \hline
        \textbf{n} & \textbf{Query} & \textbf{Time-predicate (ms)}  & \textbf{Time-rule (ms)} & \textbf{HTTP-request-rule} & \textbf{Res-rule} \\
        \hline
        100 & Q1 & x & 8,892& 3 & 0 \\
        100 & Q2 & x & 3,541& 3 & 1 \\
        100 & Q3 & x & 59,274& 8 & 8,166 \\
        \hhline{|=|=|=|=|=|=|}
        1000 & Q1 & x & 1,171& 3 & 0 \\
        1000 & Q2 & x & 734& 3 & 1 \\
        1000 & Q3 & x & 39,987& 51 & 8,166 \\
        \hline
    \end{tabular}
    \caption{
    The predicate-based (-predicate) reachability criterion is not able to execute the queries. 
    The rule-based (-rule) criterion performs better in term of execution time (Time) with a larger number of fragments even when performing more HTTP requests (HTTP-request).
    Q4 is not depicted because the instances were not able to terminate before the timeout.
    }
    \label{tab:result}
    \vspace*{-0.15cm}
\end{table}

The queries were executed using two configurations.
In the first configuration, we use a predicate-based reachability criterion where the engine follows each link of the fragmented dataset.~\sepfootnote{sf:predicateCriterion}
For the second one, we use our rule-based reachability criterion approach.
As shown in Table \ref{tab:result} no queries could be answered within the timeout by following every fragment.
A possible explanation is the high number of HTTP requests performed~\cite{Hartig2016} leading to non-relevant data sources. 
With our rule-based reachability criterion, the queries executed over the 1000 nodes fragmentation perform better than the ones with 100 nodes.
The query execution time has a percentage of reduction of 86\% with Q1 and 79\% with Q2 compared to the fragmentation with 100 nodes.
With Q3 we see that the percentage of reduction is 33\%, this lowering of performance gain might be caused by the increase by a factor of 6 in HTTP requests.
This raises an interesting observation because we do not observe a reduction in execution time with a reduction in HTTP requests.
Previous research has proposed that inefficient query plans might be the bottleneck of some queries in structured environments~\cite{taelman2023,eschauzier_quweda_2023}.
However, our results seem to show that the size of the internal triple store might have a bigger impact on performance than noted in previous studies.
As large-scale link traversal over the web will result in the acquisition of a large number of triples, a future interesting research direction would be to find ways to remove triples that are certain to not lead to a query result from the internal triple store.
The query Q4 was not able to be answered, with any setup, because the query requires a larger number of fragments than the other to be processed.

\section{Conclusion}

This paper reported on preliminary tests to add guided link traversal support into the Comunica querying engine using a rule-based reachability approach.
A similar approach could be performed with other SPARQL query engines supporting Link Traversal Query Processing. 
Our preliminary results show that our rule-based reachability criterion can significantly reduce the execution time of queries aligned with hypermedia description constraints compared to predicate-based reachability
opening the possibility for faster and more versatile traversal-based query execution over fragmented RDF documents.
Our experiment also highlights that the size of the internal data store might have more impact on performance than noted in previous studies.
In future work, we will perform more exhaustive evaluations of other types of domain-oriented fragmentation strategies such as string and geospatial evaluations,
and investigate how to generalize our approach to support more expressive online reasoning for online source selection during traversal queries.
Furthermore, we also showed there might still be room for optimization by researching ways for pruning useless triples from the internal triple store during the link traversal process.

% --- Bibliography ---

\bibliography{references}

\end{document}